\documentstyle[12pt]{article}
\renewcommand{\theequation}{\arabic{section}.\arabic{equation}}
\newcommand{\beqn}{\begin{equation}}
\newcommand{\eeqn}{\end{equation}}
\newcommand{\beqnar}{\begin{eqnarray}}
\newcommand{\eeqnar}{\end{eqnarray}}
\newcounter{abc}
\newcommand{\beqna}{\renewcommand{\theequation}{\arabic{equation}\alph{abc}}
                    \setcounter{abc}{1}
                    \begin{equation}}
\newcommand{\eeqna}{\end{equation}
                    \renewcommand{\theequation}{\arabic{equation}}}
\newcommand{\beqnb}{\renewcommand{\theequation}{\arabic{equation}\alph{abc}}
                    \setcounter{abc}{2}
                    \addtocounter{equation}{-1}
                    \begin{equation}}
\newcommand{\eeqnb}{\end{equation}
                    \renewcommand{\theequation}{\arabic{equation}}}
\newcommand{\beqnc}{\renewcommand{\theequation}{\arabic{equation}\alph{abc}}
                    \setcounter{abc}{3}
                    \addtocounter{equation}{-1}
                    \begin{equation}}
\newcommand{\eeqnc}{\end{equation}
                    \renewcommand{\theequation}{\arabic{equation}}}
\newcommand{\beqnd}{\renewcommand{\theequation}{\arabic{equation}\alph{abc}}
                    \setcounter{abc}{4}
                    \addtocounter{equation}{-1}
                    \begin{equation}}
\newcommand{\eeqnd}{\end{equation}
                    \renewcommand{\theequation}{\arabic{equation}}}
\newcommand{\beqne}{\renewcommand{\theequation}{\arabic{equation}\alph{abc}}
                    \setcounter{abc}{5}
                    \addtocounter{equation}{-1}
                    \begin{equation}}
\newcommand{\eeqne}{\end{equation}
                    \renewcommand{\theequation}{\arabic{equation}}}
\newcommand{\beqnarlett}{
                    \renewcommand{\theequation}{\arabic{equation}\alph{abc}}
                    \begin{eqnarray}}
\newcommand{\eeqnarlett}{\end{eqnarray}
                    \renewcommand{\theequation}{\arabic{equation}}}
\begin{document}
\title{Reaction-diffusion processes as physical realizations of Hecke algebras}
\author{Vladimir Rittenberg \\
Physikalisches Institut, Universit\"at Bonn, Nussallee 12, \\
53115 Bonn, Germany.}
\date{}
\maketitle
\vspace{2cm}
\begin{abstract}
The master equation describing non-equilibrium one-dimensional problems like
diffusion limited reactions can be written as a euclideen Schr\"odinger
equation in which the wave function is the probability distribution and the
Hamiltonian is that of a quantum chain with nearest-neighbour interactions.
Since many one-dimensional chains are integrable, this opens a new field
of applications. For many reactions the Hamiltonian can be written as the sum
of generators of various quotients of the Hecke algebra giving hermitian and
non-hermitian (for irreversible processes) representations.
\end{abstract}

\section{Introduction}
It is the aim of this lectures to show that mathematical notions like
associative algebras and quantum groups which have appeared in the context
of integrable systems and equilibrium statistical mechanics find a natural
application in non-equilibrium statistical mechanics also. It is well known,
for example, in the literature that several integrable quantum chains
corresponding to magnetic systems [1] can be represented as sums of generators
of Hecke algebras $H_n (q)$ when certain artificial interactions are added
in the bulk and at the surface ($q$ is the same parameter which appears in
quantum groups). We will show that the master equation describing the
dynamics of some chemical processes limited by diffusion [2] gives
representations of Hecke algebras, where the supplementary interactions
appear naturally and the deformation parameter $q$ occurring in quantum groups
gets a simple physical meaning. Moreover we would like to stress that the
quantities of interest in non-equilibrium statistical mechanics
(concentrations,
correlation functions) are very different than those occurring in equilibrium
problems and that the knowledge obtained in integrable systems to get spectra
and wave-functions of quantum chains Hamiltonians can find new fields of
applications.

This lectures are organized as follows. In Sec. 2 we review the definitions of
the Hecke algebras and remind the reader about their known hermitian
realizations ($q$-real) through the Perk-Schultz [3] quantum chains.

In Sec. 3 we describe some reaction-diffusion phenomena and show their critical
properties. For the case of one-dimensional chemistry (molecules diffuse and
react on one line - not a very realistic experimental situation) we give the
master equation and the corresponding Hamiltonian which gives the
time-evolution of the system. In Sec. 4 we consider the simple example of the
one-species annihilation model. This example is rich enough to illustrate
how integrable systems, Hecke algebras and quantum groups appear in
non-equilibrium phenomena. More examples are given in Sec. 5. The examples are
of two kinds. Some chemical reactions are given by the same quotients of the
Hecke algebra as those appearing in the one-species annihilation model which
implies that the energy levels (not the degeneracies) are the same. Other
chemical reactions give new quotients of the Hecke algebra. As a matter of
fact, we show that all the quotients of the Hecke algebra have realizations
in chemical processes.

At this point the reader might wonder if the occurrence of Hecke algebras
in reaction-diffusion processes is not just a curiosity, the answer is that
starting with a realization of the Hecke algebra one can construct through
"Baxterisation" [4] a spectral dependent matrix $\check{R} (u)$ satisfying
the Yang-
Baxter relations which is the first step towards the exact integrability of
the Hamiltonian.

These lectures are based on Refs. [5 - 7]. For applications of the ideas
presented here, see Refs. [8 - 10]. For related topics and complementary
approaches see Refs. [11 -13].

\section {The Hecke algebra and its quotients}
\setcounter{equation}{0}
The Hecke algebra $H_n (q)$ (with $n = L - 1$) is an associative algebra with
generators $e_i \; ( i = 1, \cdots , L - 1)$ satisfying the relations:
\begin{equation}
e_i e_{i \pm 1}  e_i - e_i \; = \; e_{i \pm 1} e_i e_{i \pm 1} - e_{i \pm 1}
\end{equation}
\begin{equation}
[e_i , e_j] = 0 \; , \; \mid i - j \mid \geq 2
\end{equation}
\begin{equation}
e^2_i = (q + q^{-1} ) e_i
\end{equation}
where $q$ is in general a complex parameter. In our applications $q$ is always
real.

One can define [14] a sequence of ($P,M$) quotient algebras of $H_n (q) $
($P$ and $M$ are non-negative integers). Since the definition is quite
complicated, we will give a few examples only. For the quotient (1,0), one
just take $e_i = 0$. This quotient is not relevant for our purposes. The
(2,0) quotient is the Temperley-Lieb algebra:
\begin{equation}
e_i e_{i \pm 1} e_i - e_i \; = 0
\end{equation}
The ($1, 1$) quotient reads:
\begin{equation}
(e_i e_{i + 2} ) e_{i + 1} (q + q^{-1} - e_i) (q + q^{-1} - e_{i + 2} ) = 0
\end{equation}
Different representations of the (2,0) and (1,1) quotients will be given in
Secs. 4 and 5. We will not give here the definitions of other quotients but
just mention that each ($P$, $M$) quotient has as representative one of the
($P$, $M$) Perk-Schultz [3] quantum chains where the Hamiltonian $H^{(P, M)}$
is given by the sum of $H_n (q)$ generators:
\begin{equation}
H^{(P, M)} = \sum^{L - 1}_{j = 1} \; e_j^{(P, M)}
\end{equation}
where
\begin{eqnarray}
e_j^{(P, M)} \; = \; \frac{p + p^{-1}}{2} -
\left( \sum_{\alpha \neq \beta} E^{\alpha \beta}_j \right.
E^{\beta \alpha }_{j + 1} \; + \;
\frac{q+q^{-1}}{2} \sum^{N - 1}_{\alpha = 0} \varepsilon_\alpha
E^{\alpha \alpha}_j E^{\alpha \alpha}_{j + 1} \nonumber \\
+ \frac{q - q^{-1}}{2} \sum_{\alpha \neq \beta} sign (\alpha - \beta)
\left. E^{\alpha \alpha}_j  E^{\beta \beta}_{j + 1} \right)
\end{eqnarray}
with $N = P + M$. The $N \times N$ matrices $E^{\alpha \beta}$ have elements:
\begin{equation}
\left( E^{\alpha \beta } \right)_{\gamma \delta} = \delta_{\alpha \gamma}
\delta_{\beta \delta} \; ( \alpha , \beta = 0, 1, \cdots , N - 1)
\end{equation}
and
\begin{equation}
\varepsilon_0 = \varepsilon_1 = \cdots = \varepsilon_{P - 1} = - \varepsilon_P
= \cdots = - \varepsilon_{P + M - 1} \; = 1
\end{equation}

The quantum chains $H^{(P, M)}$ are $U_q (su (P/M))$ invariant [15] and have
the important property that they contain all the irreducible representations
of the $(P, M)$ quotient. In the examples we are going to mention, if a chain
corresponds to a certain quotient $(P, M)$ then its energy levels (not
the degeneracies!) are the same as those of the representative of this
quotient given by eqs. (2.6) and (2.7). For a detailed discussion of the
degeneracies and the role of finite-dimensional representations of the affine
quantum groups $U_q (\widehat{s u (n)})$, see Ref. [7].

The generators of the ($2, 0$) and ($1, 1$) quotients can be written in terms
of the Pauli matrices
$ \sigma^x_j , \sigma^y_j$ and $\sigma^z_j (j = 1, \cdots , L)$:
\begin{eqnarray}
e^{(2, 0)}_j =
\begin{array}{c}
- \frac{1}{2}  \left[ \sigma^x_j \sigma^x_{j + 1} \; + \; \sigma^y_j
\sigma^y_{j + 1} \; + \;
\frac{q + q^{-1}}{2} \sigma^z_j \sigma^z_{j + 1}  \right. \\
- \frac{q - q^{-1}}{2} \;
\left( \sigma^z_j - \sigma^z_{j + 1} \right) \; - \;
\left. \frac{(q + q^{-1})}{2} \right]
\end{array}
\end{eqnarray}

The Hamiltonian $H^{(2, 0)}$ (see eqs. (2.6) and (2.10)) is the
$U_q (su(2))$ invariant quantum chain of Pasquier and Saleur [16]. The
$U_q (su(2))$ generators:
\begin{eqnarray}
S^z     & = & \frac{1}{2} \sum^L_{j = 1} \sigma^z_j \\
S^\pm   & = &             \sum^L_{j = 1} S^\pm_j    \\
S^\pm_j & = & q^{{\frac{1}{2}} \sum^{j - 1}_{k = 1} \sigma^z_k }
 \; \sigma^\pm_j \;
q^{{- \frac{1}{2}} \sum^L_{k = j + 1} \sigma^z_k}
\end{eqnarray}
commute with $H^{(2, 0)}$ and satisfy the relations:
\begin{eqnarray}
\left[ S^z , S^\pm \right] & = & \pm S^\pm   \\
\left[ S^+ , S^- \right] & = & [ 2 S^z ]_q \; = \;
\frac{q^{2 S^z} - q^{-2 S^z}}{q - q^{-1}}
\end{eqnarray}

These Hecke generators of the $(1, 1)$ quotient are:
\begin{equation}
e_j^{(1, 1)} = -\frac{1}{2}
\left[ \right. \sigma^x_j \sigma^x_{j + 1} \; + \; \sigma^y_j \sigma^y_{j + 1}
\; + \; q^{-1} \sigma^z_{j + 1} \;
+ \; q \sigma^z_j - (q + q^{-1}) \left. \right]
\end{equation}

The Hamiltonian $H^{(1, 1)}$ commutes [17] with the generators of the
superalgebra
$U_q (su (1 / 1)), S^z$ (see eq. (2.11)) and $T^\pm$ :
\begin{equation}
T^\pm  = q^{\frac{1 - L}{2}} \sum^L_{j = 1} \; q^{j - 1} \;
e^{\frac{i \pi}{2} \sum^{j - 1}_{k = 1} (\sigma^z_k + 1) } \; \sigma^\pm_j
\end{equation}
\begin{equation}
\left[ S^z , T^\pm \right]  = T^\pm  \qquad , \{  T^+ , T^- \} = \;
\frac{q^L - q^{-L}}{q - q^{-1}}
\end{equation}

That much for the mathematical introduction.

\section{Reaction - diffusion phenomena}
\setcounter{equation}{0}

One of the simplest examples of a reaction-diffusion process is the one
in which one has two types of molecules, say $A$ and $B$ which diffuse in a
milieu (a gel solution) and react to give an inert product which precipitates.
If at the time $t = 0$ the concentrations of the molecules $A$ and $B$ are
\begin{equation}
C_A (t = 0) = C_B (t = 0) = \varphi_A
\end{equation}
one observes that for long times, one has:
\begin{equation}
C_A (t) = C_B (t) = \frac{\lambda}{t^{d/4}}
\end{equation}
where $d$ is the space-dimensions of the milieu and $\lambda$ is a constant
independent on the initial concentration $\varphi_A$. The independence of the
late-times concentrations on the initial conditions is called self-organisaion.
Notice the power fall-off of the concentrations which implies a critical
(massless) behaviour. For the experimental setting and results see Ref. [18].
There are many theoretical and experimental questions to ask about such a
process, we are going to mention a few at the end of this Section.

In order to model the process just described in $d = 1$, we will consider
a one-dimentional chain with $L$ sites. To shorten the notations we will
denote by $A_0$ a vacancy and by $A_1$ and $A_2$ the molecules $A$ and $B$
respectively. At $t = 0$ we give a probability to find a distribution of the
molecules $A_\beta (\beta = 0, 1, 2)$ on the $L$ sites (a vacancy $A_0$ will
also be called a molecule). In order to describe the time evolution of the
system we will assume that there are only two-bodies nearest-neighbours
"interactions" decribed by the rates $\Gamma^{\alpha , \beta}_{\gamma ,
\sigma}$
which give the probability per unit time for two molecules $A_\alpha $ and
$A_\beta$ situated on the sites $i$ and $i + 1 (i = 1, \cdots , L - 1$ to
change into the molecules $A_\gamma $ and $A_\sigma$. Notice that in general
the processes don't have to be left-right symmetric
$ \left( \Gamma^{\alpha , \beta}_{\gamma , \sigma} \neq \right.
\left. \Gamma^{\beta , \alpha}_{\sigma , \gamma} \right)$.
For the process described above one can take only a few non-zero rates:
\begin{eqnarray}
A_k + A_0 \to A_0 + A_k  & {\mbox {(diffusion to the right)}} \quad
 (k = 1 , 2) \\
A_0 + A_k \to A_k + A_0  & {\mbox {(diffusion to the left)}} \quad
(k = 1 , 2) \\
A_1 + A_2 \to A_0 + A_0  & {\mbox {(reaction)}} \\
A_2 + A_1 \to A_0 + A_0  & {\mbox {(reaction)}}
\end{eqnarray}

with only two independent rates $D$ and $\Gamma$.
\begin{equation}
\Gamma^{k , 0}_{0 , k } = \Gamma^{0 , k}_{k , 0} =
D \quad  (k = 1 , 2 ) \; , \;
\Gamma^{1 , 2}_{0 , 0 } = \Gamma^{2 , 1}_{0 , 0} = \Gamma
\end{equation}

The time evolution of the system is given by a master equation which gives the
probability to find a certain configuration of the molecules $A_\beta$ at a
time $t$ on the $L$ sites for a given probability distribution at $t = 0$.
Knowing the probability distribution at $t$, one can compute various average
quantities like the concentrations for $L$ sites and then take the large
$L$ limit in order to get the result given by eq. (3.2). At this point we
will consider the problem in its full generality.

Let us consider an open chain with $L$ sites, where at each site
$i = 1 , 2, \cdots, L$ we attach a variable $\beta$ (representing the
molecule $A_\beta$) taking $N$ integer values
$(\beta = 0, 1 , \cdots, N - 1)$. By convention we attach the value
$\beta = 0$ to a vacancy (inert state). The master equation describing the
time evolution of the probability destribution
\begin{equation}
P (\{ \beta \} ; t) = P (\beta_{1} , \beta_2 , \cdots , \beta_L ; t)
\end{equation}
is:
\begin{eqnarray}
\frac{d P (\{ \beta\} , t)}{d t} & = &
\sum^{L - 1}_{j = 1} \left( - w_{0,0} (\beta_j , \beta_{j + 1}) \right.
P (\{ \beta\} ; t) + \nonumber \\
& + &
\sum^{N - 1}_{l, m = 0}\,^\prime w_{l , m} (\beta_j , \beta_{j + 1})
P \left( \beta_{1, \cdots ,} [ \beta_{j + l} ]_N \; , \right.
\left. [ \beta_{j + 1} + m ]_N , \cdots , \beta_L ; t) \right)
\end{eqnarray}
where the $w_{l , m}$ are related to the transition rates
$\Gamma^{\alpha , \beta}_{\gamma , \sigma}$ through the relations:
\begin{equation}
\Gamma^{\alpha , \beta}_{\gamma , \sigma} \; = \;
w_{\gamma - \alpha , \sigma -\beta } ( \alpha , \beta)
\end{equation}
In eq. (3.9) the prime in the second sum indicates the exclusion of the pair
$l = m = 0$ and the symbol $ [x + y]_N$ means the addition $(x + y)$ modulo
$N$.
The rate $w_{0 0} (\alpha , \beta )$ gives the probability per unit of time
that the configuration of two neighbouring sites $(\alpha , \beta )$ is
changed. Using eq. (3.10), the conservation of probabilities implies:
\begin{equation}
w_{0 , 0} (\alpha, \beta ) = \sum_{r , s}^\prime
w_{r , s} (\alpha - r , \beta - s)
\end{equation}
where $r = s = 0$ is again excluded. The notations used in eq. (3.9) are
convenient to read symmetries. The rates are always non-negative and symmetries
imply that either some rates are equal or that they vanish. For example to
have $Z_N$ symmetry implies that all rates $w_{l,m} (\alpha , \beta )$
vanish except those satisfying the relation:
\begin{equation}
[ l + m ]_N = 0
\end{equation}
Parity conservation (left - right symmetry) implies:
\begin{equation}
w_{l , m} ( \alpha , \beta ) = w_{m , l} ( \beta , \alpha )
\end{equation}

The master equation (3.9) can be interpreted as a Schr\"odinger equation
\begin{equation}
\frac{d}{d t} \mid P > \; = \; - H \mid P >
\end{equation}
in Euclidean time if we identify the probability distribution
$P ( \{ \beta \} ; t)$ as the wave function. The Hamiltonian in eq. (3.14)
\begin{equation}
H = \sum^{L - 1}_{j = 1} H_j
\end{equation}
acts in a Hilbert space of dimension $N^L$ while $H_j$ acts in the subspace
$V^{( j )} \otimes V^{( j + 1 )}$ and is given by
\begin{equation}
H_j = U_j - T_j
\end{equation}
where
\begin{eqnarray}
T_j & = & \sum^{N - 1}_{l , m = 0}\! ^\prime \;
\sum^{N - 1 }_{\alpha , \beta = 0} \;
w_{l , m} (\alpha , \beta ) E^{\alpha , [ \alpha + l ]_{N}} \; \otimes
E^{\beta , [ \beta + m ]{_N}} \nonumber \\
U_j & = & \sum^{N - 1 }_{\alpha , \beta = 0} \; w_{0 , 0} (\alpha , \beta) \;
E^{\alpha \alpha} \otimes E^{\beta \beta}
\end{eqnarray}
The matrices $E^{\alpha \beta }$ have been defined in eq. (2.8). We now show
how to compute physical quantities from the knowledge of $\mid P >$.
We define a basis
\begin{equation}
\mid \{ \beta \} > = \mid \beta_1 , \beta_2 , \cdots , \beta_L > \; , \;
< \{ \beta^\prime \} \mid \{ \beta \} > =
\delta_{ \{ \beta \} , \{ \beta^\prime \}}
\end{equation}
and is this basis the ket-vector $\mid P >$ is:
\begin{equation}
\mid P > = \sum_{\{  \beta \}} P ( \{ \beta \}, t ) \mid \{ \beta \} >
\end{equation}
to the initial probability distribution $P ( \{ \beta \} , t = 0 )$
corresponds the ket-vector $\mid P_0 >$.

We define a left-vacuum bra-state:
\begin{equation}
< 0 \mid = \sum_{ \{ \beta \}} < \{  \beta \} \mid
\end{equation}
then the conservation of probabilities (3.11) implies
\begin{equation}
< 0 \mid H = 0
\end{equation}

If $X ( \{ \beta \} )$ is an observable, then its average at a time $t$ is
\begin{equation}
< X >_t = \sum_{ \{ \beta \}} X ( \{ \beta \} )  P (\{ \beta \} ; t ) =
< 0 \mid X e^{- H t} \mid P_0 >
\end{equation}
Notice that the expression (3.22) is very different of the vacuum
expectation values encountered in equilibrium statistical mechanics. Here
the initial probability distribution $ \mid P_0 > $ appears in an essential
way and can introduce new lenghts scales into the problem.

If the energy levels $E_\lambda$ and the eigen-functions $\mid \psi_\lambda >$
are known
\begin{equation}
H \mid \psi_\lambda > = E_\lambda \mid \psi_\lambda >
\end{equation}
with
\begin{equation}
\mid P_0 > = \sum_\lambda  a_\lambda \mid \psi_\lambda >
\end{equation}
$< X >_t$ can be rewritten as:
\begin{equation}
< X >_t = \sum_\lambda a_\lambda e^{- E_{\lambda} t} < 0 \mid X \mid
\psi_\lambda >
\end{equation}
The system has a critical behaviour if in the thermodynamic limit
$( L \to \infty)$ all the $E_\lambda $ vanish (keep in mind that because of
eq. (3.21), the ground state has always energy zero). There is a profound
difference between a massless regime often (but not always!) seen in
equilibrium statistical mechanics when
\begin{equation}
{{\lim} \atop {L \to \infty }} \; L E_\lambda = {\mbox{constant}}
\end{equation}
and one has conformal invariance and the present case when
\begin{equation}
{{\lim} \atop { L \to \infty}} \; L^2 E_\lambda = {\mbox{constant}}
\end{equation}
This can be easily understood since the diffusion equation looks like a non-
relativistic Schr\"odinger equation and eq. (3.27) reflects a non-relativistic
(quadratic) dispersion relation. Quadratic dispersion relations are met also
in equilibrium problems like in the Pokrovsky-Talapov [19] phase transition.

Before closing this Section, let us observe that it is often very useful to
perform similarity transformations in the master equation (3.9). This implies
considering instead of  $P ( \{ \beta \} ; t )$, another function
$\psi ( \{ \beta \}, t )$ related to the first through the relation:
\begin{equation}
P (\{ \beta \} ; t) = \Phi \psi ( \{ \beta \} , t)
\end{equation}
where $\Phi$ acts in the $\{ \beta \}$ space but is time independent. For our
purposes here, we take $\Phi$ diagonal and of a factorized form:
\begin{equation}
\Phi = \prod^L_{k = 1} h^{(k)} ( \beta_k )
\end{equation}
The new master equaiton reads:
\begin{eqnarray}
\frac{d \psi ( \{ \beta \} ; t )}{d t} = &
\sum^{L - 1}_{j = 1} ( - w_{0 , 0} (\beta_j , \beta_{j + 1}) \psi
(\{ \beta \} , t) + \qquad \nonumber \\
& + \sum^{N - 1{\prime}}_{l , m = 0} W_{l , m} (\beta_j , \beta_{j + 1})
\psi \;
( \beta_{1, \cdots ,} [ \beta_j + l]_N , [\beta_{j + 1} + m]_N , \cdots ,
\left. \beta_L ; t ) \right)
\end{eqnarray}
where
\begin{equation}
W^{(j)}_{l , m } (\beta_j , \beta_{j + 1}) =
\frac{h^{(j)} (\alpha_{j + l})}{h^j (\alpha_j )} \cdot
\frac{h^{(j + 1 )} ( \alpha_{j + 1} + m)}{h^{(j + 1)} ( \alpha_{j + 1})}
w_{l , m } (\beta_j , \beta_{j + 1})
\end{equation}
Accordingly, after the similarity transformation, the new Hamiltonian
(3.15) - (3.16), has the $w_{l m} (\alpha , \beta)$ in the expression of the
$T_j$'s replaced by the $W^{(j)}_{e , m}$'s given by eq. (3.31).

\section{The annihilation-diffusion process}
\setcounter{equation}{0}

We now present a simple example in which a single spacies of molecules
$A_1$ annihilate:
\begin{equation}
A_1 + A_1 \to A_0 + A_0
\end{equation}
with a rate $w_{1, 1} ( 0,0) = a$, diffuse to the right
\begin{equation}
A_1 + A_0 \to A_0 + A_1
\end{equation}
with a rate $w_{1 1} (0,1) = D_R$, and diffuse to the left
\begin{equation}
A_0 + A_1 \to A_1 + A_0
\end{equation}
with a rate $w_{1 1} (1,0) = D_L$. No other processes are considered. We
will denote
\begin{equation}
D = \sqrt{D_L D_R} \; , \; \; q = \sqrt{\frac{D_L}{D_R}}
\end{equation}
and make a similarity transformation (see eq. (3.29)):
\begin{equation}
\frac{h^{(k)} (1)}{h^{(k)} (0)} = \frac{q^{1 - k}}{\lambda}
\end{equation}
where $\lambda$ is a free parameter. We also choose the time units such that
$D = 1$ such that the whole process depends on two physical parameters $a$
and $q$. The Hamiltonian describing this process is:
\begin{equation}
H = H_0 + H_1
\end{equation}
where
\[
H_0 = \; \;  -\frac{1}{2} \sum^{L - 1}_{j = 1} \left[ \sigma^x_j \sigma^x_{j +
1} \right.
+ \sigma^y_j \sigma^y_{j + 1} + \Delta \sigma^z_j \sigma^z_{j + 1 } +
(1 - \Delta^\prime ) (\sigma^z_j + \sigma^z_{j + 1})
\]
\begin{equation}
- \frac{1}{2} (q - q^{-1}) (\sigma^z_j - \sigma^z_{j + 1}) + 2 \Delta^\prime
\left. - \Delta - 2 \right]
\end{equation}
\begin{equation}
H_1 = - \frac{a}{\lambda^2} \sum^{L - 1}_{j = 1} q^{1 - 2j} \sigma^+_j
\sigma^+_{j + 1}
\end{equation}
where
\begin{equation}
\Delta = \frac{q + q^{-1}}{2} - \frac{a}{2} \; , \; \; \Delta^\prime = 1 -
\frac{a}{2} \; \; ( \Delta^\prime  < \Delta )
\end{equation}
Notice that $H_0$ is hermitian and that the non-hermitian $H_1$ has a
coupling constant proportional to $\lambda^{-2}$ and that the parameter
$\lambda $
is given by the similarity transformation (4.5). Since the spectrum of the
Hamiltonian $H$ can't depend on a similarity transformation this implies that
the spectrum of $H$ coincides with the spectrum of $H_0$. The spectrum of $H_0$
is known since this is the $XXZ$ chain in an external $Z$ field [20] where
the Bethe-Ansatz applies. Leaving asside the surface fields, $H_0$ can be
written as
\begin{equation}
H^\prime_0 = - \frac{1}{2} \sum^{L - 1}_{J = 1} \left[ \sigma^x_j \right.
\sigma^x_{j + 1} + \sigma^y_j \sigma^y_{j + 1} + \Delta \sigma^z_j
\left. \sigma^z_{j + 1} + h (\sigma^z_j + \sigma^z_{j + 1}) \right]
\end{equation}
In the thermodynamic limit $( L \to \infty )$ the spectrum of $H_0$ is
massive if $h = 1 - \Delta^\prime > 1 - \Delta $. This implies that for
assymmetric diffusion $(q \neq 1)$ in an open chain the spectrum is always
massive. If $h = 1 - \Delta $ (which implies $q = 1$) the spectrum is massless
and corresponds to a Pokrovsky-Talapov phase transition with a quadratic
dispersion relation.

It is now time to make the connexion between the annihilation-diffusion process
and the Hecke algebra. Let us first consider pure diffusion
$( a = 0)$. It is easy to check that in this case (see eq. (2.10)) that
\begin{equation}
H = H_0 = \sum^{L - 1}_{j = 1} e^{(2 , 0)}_j
\end{equation}
This result is very interesting since it gives a simple physical interpretation
to the parameter $q$ of the quantum group $U_q (su (2))$. It is (see eq.
(4.4)) the square root of the ratio of the left and right diffusion constants.
The spectrum is of course massive if $q \neq 1$. This is true for the open
chain. If one considers the assymmetric diffusion process on a ring however
(periodic boundary conditions), the similarity transformation (4.5)
induces a boundary term with a strength of the order of the volume which
makes the system massless [21].

Let us now take
\begin{equation}
a = q + q^{-1}
\end{equation}
which corresponds to the physical situation if two molecules $A_1$ are
on neighbouring sites, they annihilate with probability one. In this case
\begin{equation}
H_0 = \sum^{L - 1}_{j = 1} \; e^{(1, 1)}_j
\end{equation}
moreover
\begin{equation}
H = \sum^{L - 1}_{j = 1} \; f_j
\end{equation}
where the non-hermitian matrices $f_j$ satisfy the relations (2.1 - 3)
and (2.5) giving
a representation of the $(1,1)$ quotient of the Hecke algebra.

The example considered in this Section illustrates the general structure which
is going to be described in the next section. Pure diffusion leads to
quotients of the type $(N, 0)$ corresponding to $U_q (su (N, 0))$ symmetric
quantum chains, non-reversible processes give $(M, P) (P \neq 0)$ quotients.

\section{Various representations of the Hecke algebra given by
reaction-diffusion processes}
\setcounter{equation}{0}

(a) {\em Diffusion processes.} We consider $(N - 1)$ types of molecules
($A_1 , \cdots A_{N - 1}$) which diffuse to the right:
\begin{equation}
A_b + A_0 \to A_0 + A_b \; (b = 1 , \cdots , N - 1)
\end{equation}
with equal rates $\Gamma_R$ and to the left:
\begin{equation}
A_0 + A_b \to A_b + A_0 \; (b = 1, \cdots , N - 1),
\end{equation}
with equal rates $\Gamma_L$. Choosing the time scale in such a way that
the diffusion constant \\
$D = \sqrt{\Gamma_L \Gamma_R} = 1$ and introducing
$q = \sqrt{\Gamma_R / \Gamma_L}$, up to an equivalence transformation, we
obtain $H_j = e_j$ corresponding to the quotient $(2,0)$.

(b) {\em Diffusion and interchange processes.}
To the diffusion processes (5.1) and (5.2) we add the interchange to the right
processes:
\begin{equation}
A_b + A_c \to A_c + A_b \; (b \neq c , b > c ),
\end{equation}
with rates $\Gamma_R$ and with rates $\Gamma_L$ if $b < c$. In this case the
quotient is (N, 0) and the obtained representation of $H_n (q)$ is hermitian
(the same applies to the diffusion processes of case (a).

(c) {\em Diffusion, interchange and coagulation processes.}
We now add to the processes (5.1 - 3) only one reaction for the molecule
$A_1$ (the molecules are ordered!):
\begin{equation}
A_1 + A_1 \to A_0 + A_1,
\end{equation}
with a rate $\Gamma_R$ and the process where $A_0$ and $A_1$ interchange
their position in the final state with rate $\Gamma_L$. In this case the
quotient is $(N - 1 , 1) $ and the representation of $H_n$ is non-hermitian.
As a rule, non-hermitian representations occur for quotients $(P, M)$ with
$M \neq 0$ i. e. when the representative Hamiltonians $H^{(P, M)}$
are supersymmetric.

One obtains new quotients of $H_n$ generalizing the reaction (5.4) by taking
\begin{equation}
A_r + A_r \to A_s + A_r \; \; ({\mbox{rate}} \Gamma_R),
\end{equation}
\begin{equation}
A_{r^\prime} + A_{r^\prime} \to A_{s^\prime} + A_{r^\prime}
\; \; ({\mbox{rate}} \Gamma_R),
\end{equation}
where $s = r \pm 1 , s^\prime = r^\prime \pm 1$, and
$ r \neq r^\prime , s^\prime ; s \neq r^\prime , s^\prime $. In this way one
obtains the quotient $(N - 2,2)$. Taking more reactions like in eqs.
(5.5), (5.6) one can get arbitrary quotients $(P,M)$.

(d) {\em Diffusion-polymerisation (modulo N) processes.} In this case besides
the diffusion (eqs. 5.1), (5.2), one considers the processes
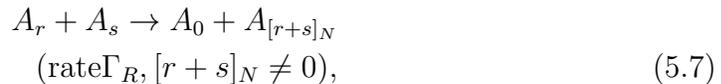
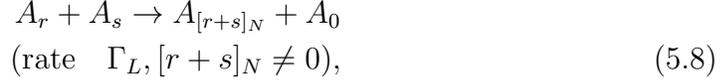
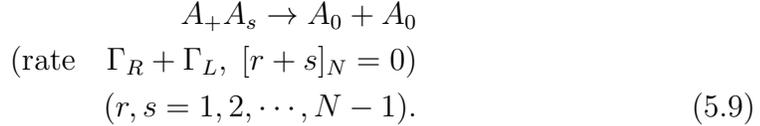
\begin{eqnarray}
A_r + A_s \to A_0 + A_{[r + s]{_N}} \nonumber \\
{\mbox{(rate} } \Gamma_R , [r + s]_N \neq 0 ),
\end{eqnarray}
\begin{eqnarray}
A_r + A_s \to A_{[r + s]{_N}} + A_0 \nonumber \\
{\mbox{(rate}} \quad \Gamma_L , [r + s]_N \neq 0),
\end{eqnarray}
\begin{eqnarray}
A_ + A_s \to A_0 + A_0 \nonumber \\
{\mbox{(rate}} \quad \Gamma_R + \Gamma_L , \; [r + s]_N = 0 ) \nonumber \\
( r, s = 1, 2, \cdots, N - 1).
\end{eqnarray}
For all these processes one obtains the quotients $(1, 1).$

(e) {\em Diffusion, interchange and annihilation processes.}
Because of interchange processes, the labels of the molecules are ordered;
we can now add the following two processes:
\begin{equation}
A_m + A_m \to A_{m + 1} + A_{m + 1} \; \; ({\mbox{rate}} \Lambda_1 ),
\end{equation}
\begin{equation}
A_m + A_m \to A_{m - 1} + A_{m - 1} \; \; ({\mbox{rate}} \Lambda_{-1} ),
\end{equation}
with $\Lambda_1 + \Lambda_{-1} = \Gamma_R + \Gamma_L $. In this way we
obtain again the quotient $(2, 1)$.

The reader might wonder how the above $H_n (q)$ quotients were obtained.
{}From the work done in ref. [5] we have developed a certain intuition for
chemical processes up to $N = 3$. We have then used the computer to check
that no other solutions exist for $N = 2$ and $3$ and generalized the
solutions obtained for $N$ up to $3$ to all $N$ and checked them. There are
certainly other solutions which one can discover studying sytems with
$N = 4$ and certainly examples of the quotients $(2, 2)$ different from the one
given in eqs. (5.5), (5.6) will show up. We hope that the present effort
to connect representatives of $H_n (q)$ and chemical processes will help to
compute various correlation functions for the latter. One immediate consequence
of the above mentioned connection is that all processes with $q \neq 1$ are
massive (the concentrations for example have an exponential fall-off in time)
and all those with $q = 1 $ are massless (algebraic fall-off).

\end{document}